# AI SciBrief as a Gateway to Research: A Framework for Onboarding Students into New Research Areas

Andrei Lazarev
Department of Control and Applied Mathematics
Moscow Institute of Physics and Technology
Moscow, Russia
ORCID:0009-0008-1519-5820

Dmitrii Sedov
Moscow Institute of Physics and Technology
Moscow, Russia
ORCID:0009-0009-0964-7576

*Abstract*—**Students at all levels of higher education face a significant barrier in the form of information overload, which often paralyzes the initial stages of the research process and suppresses motivation. In response, this article introduces a pedagogical framework that leverages AI SciBrief, a platform powered by a Large Language Model (LLM) designed to automatically generate digests of scientific trends. We describe how this multidisciplinary tool - with initial coverage in finance, medicine, and education - can be integrated into the curriculum to overcome this "entry barrier." The framework provides concrete methodologies for utilizing these digests to facilitate topic selection for term papers, accelerate literature reviews for dissertations, and enable postgraduate students to continuously monitor emerging trends. We conclude that AI SciBrief functions as a "gateway to research" effectively reducing students' cognitive load and empowering them to transition more rapidly from information searching to knowledge creation.**

## I. Introduction

In the modern academic landscape, researchers are navigating an unprecedented "information overload". The volume of scholarly output has surged exponentially, with global scientific publications reaching 3.3 million articles in 2022 alone [1][2]. According to [3] a near-constant flood of new data, with estimates suggesting that roughly 10,000 new scientific papers are published every single day. This relentless growth in publications, which has been documented to be exponential at a rate of about 5.6% per year, creates a formidable challenge for anyone attempting to stay abreast of the latest developments in their field.

This constant flood of information poses a significant barrier, particularly for students and early-career researchers. This can lead to reduced attention spans, difficulty in focusing on complex academic tasks, and increased levels of stress and anxiety [4]. Novice researchers, in particular, struggle with challenges such as a lack of confidence in selecting a topic, difficulties in finding relevant literature among vast and sometimes misleading online data, and inadequate assistance in navigating the research process.

**For undergraduate students**, the primary obstacle is the paralyzing fear of the "blank page". Tasked with writing a term paper or a bachelor's thesis, they are often the most vulnerable to the sheer volume of available information [5]. Lacking deep subject matter expertise, they struggle to identify a research gap or even formulate a relevant and sufficiently narrow topic. The process of navigating countless databases and search results without a clear direction becomes a source of anxiety, often leading to procrastination or the selection of overly broad and unoriginal subjects. For them, the challenge is not just finding information, but finding a starting point in the first place.

**At the master's level**, the challenge shifts from topic selection to synthesis and scope. Master's candidates are expected to produce a comprehensive literature review for their dissertations, demonstrating a thorough understanding of the existing body of work. This requires them to read through hundreds of articles to identify creditable works, current debates, and methodological approaches. It demands an exhaustive and time-consuming effort to read, analyze, and synthesize a vast amount of material, which can delay the start of their original research and become a bottleneck in the timely completion of their degree.

**For doctoral candidates (Ph.D. students)**, the barrier is relevance and continuous awareness. Operating at the cutting edge of their fields, they face the relentless pressure to stay constantly informed of the latest developments, not only within their specific niche but also in adjacent and intersecting disciplines. Missing a key publication can mean that their research is accidentally outdated or fails to build on the most recent advancements. This necessity for ongoing, real-time monitoring of scientific trends is a significant cognitive load, diverting valuable time and energy that could otherwise be dedicated to their own experimental work and analysis.

The primary aim of this paper is to introduce "AI SciBrief" service not merely as a technological product, but as a pedagogical approach designed to sort out the research entry barriers faced by students. While the platform is powered by a sophisticated Large Language Model (LLM), our focus here is not on its technical architecture. Instead, we present AI SciBrief as an educational intervention aimed directly at the problem of information overload and the "blank page" paralysis that interferes with the initial stages of academic inquiry.

This article proposes a practical pedagogical approach for integrating AI SciBrief into the process of creating new papers by students. This service offers educators a structured approach to guide students at different academic levels: from undergraduates selecting their first research topic to doctoral candidates navigating the limits of their fields. By outlining specific use cases and methodologies, we aim to demonstrate how this tool can be strategically employed to lower cognitive load, cultivate research independence, and accelerate the transition from information consumption to knowledge creation. Ultimately, our goal is to reposition AI-driven tools like SciBrief from being simple information-retrieval utilities to becoming integral components of modern research-based learning.

## II. The AI SciBrief Platform: An Overview

AI SciBrief is an automated platform designed to serve as a bridge between the vast world of academic publications and the focused needs of students and researchers. It operates not as a search engine, but as a synthesis tool that condenses the latest scientific developments into an accessible format.

The platform is built on two core components. Its data foundation is the OpenAlex database, a comprehensive and open-source index of hundreds of millions of scholarly documents, ensuring that the analysis is both broad and current. For the analytical process, AI SciBrief leverages a powerful Large Language Model (LLM), specifically Google Gemini, to read, interpret, and summarize findings from thousands of new papers. The detailed technical process, from data acquisition and JSON construction to the automated generation of PDF reports, was first detailed in our foundational work on this topic [6].

The final output for the end-user is a series of professionally designed monthly PDF digests. Each digest distills the most significant trends, influential papers, and emerging concepts within a specific academic field from the preceding month. While the initial proof-of-concept and its benefits for simplifying research were demonstrated with a focus on the financial domain [7], the platform has since evolved to be intentionally multidisciplinary to underscore the versatility of its pedagogical approach. At present, AI SciBrief produces digests for the fields of Finance, Medicine, and Education. Further details and digest examples can be found on the project's official website: https://sci-brief.com

## III. Approach for Pedagogical Integration

To bridge the gap between the potential of AI SciBrief and its practical application, we propose a three-tiered pedagogical framework grounded in the established learning science concept of "scaffolding." First introduced by Wood, Bruner, and Ross (1976) [8], scaffolding refers to the temporary support provided to a learner to help them accomplish a task they could not complete on their own. This support is designed to be gradually withdrawn as the learner's competence grows. In a modern context, digital tools can serve as powerful scaffolds, particularly in complex, computer-based learning environments where they can support self-regulated learning and metacognition [9].

Our framework positions AI SciBrief as precisely such a digital scaffold. It is designed to structure the complex task of research, making key aspects of the process more visible and manageable for students engaged in inquiry-based learning [10]. The framework aligns with the evolving needs of students as they progress through their academic journey, with the "scaffold" adapting its function at each level.

### A. *Level 1: Topic Discovery (For Undergraduate Students)*

- *The Primary Challenge:* Undergraduates often face the "blank page" problem, lacking the contextual knowledge to identify current or significant research topics.
- *AI SciBrief as a Scaffold (The Method):* AI SciBrief acts as a structuring scaffold. It simplifies the initial, overwhelming task of "finding a topic" into a more manageable process of analyzing a curated digest. By presenting a concise summary of current trends, the tool makes the landscape of potential topics visible and accessible, a key function of scaffolding in inquiry learning [10]. The student's task is transformed from an open-ended search into a focused analysis, empowering them to formulate relevant research questions.
- *Key Pedagogical Outcome:* This process builds confidence and ensures research is grounded in contemporary issues. The scaffold provides a clear starting point, after which it can be "removed" as the student delves deeper into the primary literature for their chosen topic.

### B. *Level 2: Landscape Mapping (For Master's Students)*

- *The Primary Challenge:* Master's students must conduct a comprehensive literature review, which requires sophisticated skills in analysis and synthesis to identify a niche for their own contribution.
- *AI SciBrief as a Scaffold (The Method):* At this level, the tool serves as a metacognitive scaffold, supporting the development of self-regulated learning skills [9]. By analyzing a series of digests, students are not just consuming information; they are actively learning how to map a research landscape. They practice identifying key authors, debates, and research gaps, essentially using the digest as a model for the kind of synthesis they will eventually need to produce on their own.
- *Key Pedagogical Outcome:* This accelerates the literature review process and fosters strategic thinking. The scaffold doesn't write the review for the student; it provides the structure and data needed for them to build their own comprehensive understanding and position their work more effectively.

### C. *Level 3: Continuous Scanning & Interdisciplinary Synthesis (For Doctoral Candidates)*

- *The Primary Challenge:* Doctoral candidates must maintain expertise in a rapidly evolving field while also seeking innovative, interdisciplinary connections.
- *AI SciBrief as a Scaffold (The Method):* For these advanced learners, AI SciBrief functions as a conceptual scaffold. As defined in the original work by Wood et al. (1976) [8], a key role of the tutor (or in this case, the tool) is to reduce the "degrees of freedom" in a task. By monitoring both core and adjacent fields, the tool automatically filters the vast information stream, allowing the candidate to focus their cognitive resources on the highest-level task: synthesis and the creation of new knowledge.
- *Key Pedagogical Outcome:* This fosters innovation and scholarly leadership. By lowering the barrier to exploring adjacent fields, the scaffold encourages the kind of interdisciplinary thinking that leads to breakthrough research, helping the candidate to build a bridge between disciplines.

To summarize, our framework operationalizes the theory of scaffolding in a digital context, adapting the nature of the support to the learner's evolving expertise on **Table 1**:

TABLE I. THE AI SCIBRIEF SCAFFOLDING FRAMEWORK

| Academic Level | AI SciBrief as a Scaffold | Key Pedagogical Principle | Learning Outcome |
|---|---|---|---|
| **Undergraduate** | Structuring Scaffold: Makes the task of topic discovery visible and manageable | Structuring the task to reduce cognitive load (Hmelo-Silver et al., 2007) [10]. | Confidence & Relevance |
| **Master's** | Metacognitive Scaffold: Models the process of mapping a research landscape | Supporting self-regulated learning and analysis skills (Azevedo & Hadwin, 2005) [9] | Strategic Efficiency & Originality |
| **Doctoral** | Scaffold: Reduces information overload to enable high-level synthesis | Reducing degrees of freedom to allow focus on innovation (Wood et al., 1976) [8] | Innovation & Interdisciplinarity |

This framework demonstrates that AI SciBrief is not a replacement for scholarly skill, but a powerful accelerator that empowers students at each stage of their development. In the following section, we will discuss the broader implications and potential limitations of this approach.

## IV. ILLUSTRATIVE CASE STUDY: STRATEGIC TOPIC SELECTION FOR DOCTORAL DISSERTATION

To illustrate the practical application of the advanced approach, consider a hypothetical yet realistic scenario. A first-year Ph.D. student in Finance, "Alex Newman" is tasked with identifying a new and impactful research gap for his dissertation. The field is vast, and he is struggling to move beyond well-established topics.

**Step 1.** Data Collection: Following the approach, Alex gathers the last 4-6 monthly AI SciBrief digests for the "Finance" domain. This collection represents a bunch of synthesized research trends.

**Step 2.** LLM-Powered Meta-Analysis: He uploads the PDF files into a multi-document analysis LLM and prompts it with a strategic query: "Analyze these monthly finance research digests. First, identify the 3-5 most persistent research trends that appear consistently throughout the year. Second, identify any emerging intersections between these major trends that the reports highlight. Finally, point out what topics are conspicuously underrepresented or absent, suggesting potential research gaps."

**Step 3.** Synthesized Insight: The LLM processes Financial Digests and returns a powerful synthesis. It identifies following results for 4 months from April to July 2025:

- *The Widespread Integration of Artificial Intelligence and Machine Learning* - there's a growing focus on the use of Large Language Models for tasks like analyzing market sentiment from news and earnings calls. However, a critical area of investigation is the challenge of "information overload", where excessive data can paradoxically lead to poorer performance in AI models.
- *The Rise of Sustainable Finance and ESG Investing* - Researches in this area is examining the impact of ESG (Environmental, social, and governance) ratings on investor decisions, the role of green bonds, and the financial implications of climate risks.The push for sustainable finance is also driving innovation in green technologies and more transparent corporate reporting
- *The Transformation Driven by FinTech and Digitalization* - Researchers are exploring the impact of FinTech on financial inclusion, offering access to banking and investment opportunities for underserved populations. The rise of digital wallets, real-time payments, and robo-advisors are also key areas of investigations
- *A Renewed Focus on Financial Literacy and Behavioral Finance* - Studies are highlighting a significant gap in financial preparedness among adults, leading to increased financial stress. This has spurred research into more effective financial education programs and the psychological aspects of financial decision-making, a field known as behavioral finance.

Intersections:

- *AI and ESG*: A significant intersection is emerging between AI and ESG, where AI is being used to analyze vast amounts of data to generate more accurate ESG scores and identify sustainable investment opportunities.
- *FinTech and Financial Inclusion*: The synergy between FinTech and financial inclusion is a prominent area of research, exploring how digital platforms can democratize access to financial services.

**Step 4.** Outcome: Based on this LLM-driven analysis, Alex's Conclusions:

The analysis of the AI SciBrief digests provided a clear map of the financial research landscape, revealing that modern finance is dominated by four key trends: AI, ESG, FinTech, and Behavioral Finance.

Alex concluded that the most impactful research lies not within any single trend, but at their intersection, particularly between AI and ESG. The digests highlight a critical research gap here: the risk of algorithmic bias in AI models used for ESG scoring, a problem exacerbated by the lack of standardized metrics.

This has defined his dissertation topic: "*Mitigating Algorithmic Bias in Large Language Models for ESG Analysis*". This focus is timely and addresses a significant, under-researched problem.

**Step 5.** Epilogue: The Journey of Alex Newman

The journey of Alex, our hypothetical doctoral candidate, is designed to serve as a practical illustration of the pedagogical framework in action. While Alex himself is an illustrative construct, the process he followed and the conclusion he reached are both realistic and deeply relevant to the challenges facing researchers today.

His choice of dissertation topic - "Mitigating Algorithmic Bias in Large Language Models for ESG Analysis" - is not arbitrary. It is the logical and justified outcome of a structured, data-driven methodology. He did not simply choose a trending topic; he performed a longitudinal analysis of synthesized research digests to identify the most persistent forces shaping modern finance (AI and ESG) and, critically, to pinpoint a significant research gap at their intersection. His chosen topic is powerful because it is timely, addresses a tangible real-world problem where technology, ethics, and finance collide, and is grounded in a verifiable understanding of the current research landscape.

## V. Discussion and Limitations

The framework presented in this paper positions AI SciBrief not merely as a technological artifact, but as a carefully designed pedagogical scaffold. Our discussion will first contextualize the implications of this approach within the broader conversation on learning science and AI in education, before acknowledging the inherent limitations of our tool and methodology.

### A. *Discussion: Augmenting the Researcher, Not Replacing Them*

The central premise of our framework is that tools like AI SciBrief can function as a powerful form of cognitive scaffolding, a concept rooted in the foundational work of Wood, Bruner, and Ross (1976) [8]. Their original theory posits that effective support enables a learner to solve a problem or achieve a goal which would be beyond their unassisted efforts. Our framework operationalizes this by using AI to structure the overwhelming task of literature review, thereby reducing the "degrees of freedom" and allowing students to focus on the higher-order tasks of analysis and synthesis.

Crucially, our approach aligns with modern interpretations of scaffolding in digital learning environments. As argued by Azevedo & Hadwin (2005) [9], digital scaffolds should be designed to support self-regulated learning and metacognition. AI SciBrief is not a "magic box" that provides answers; it is a tool that models the process of identifying trends, thereby helping students learn how to navigate a research landscape. It serves to support, rather than bypass, the core skills required in inquiry-based learning [10].

Furthermore, our work contributes a concrete example to the ongoing discussion about the positive potential of Large Language Models in education. While much of the discourse has focused on risks, our framework responds to the call to explore the beneficial applications of generative AI in teaching and learning (Baidoo-Anu & Ansah, 2023) [12]. It represents a practical application that addresses one of the key research priorities identified in the seminal Nature commentary on ChatGPT: understanding how LLMs can enhance the scientific process itself (van Dis et al., 2023) [13]. In essence, AI SciBrief is an example of "AI for good," demonstrating a model of human-AI collaboration that empowers students rather than diminishing their role.

### B. *Limitations and Future Directions*

While we are optimistic about the potential of this framework, we recognize several key limitations that offer important directions for future research:

- *Dependence on Data Sources and Potential for Bias*: The output of AI SciBrief is fundamentally dependent on the scope and quality of its underlying data source, OpenAlex. Any inherent biases or gaps in this database will inevitably be reflected in the generated digests. Furthermore, the LLMs themselves can perpetuate or even amplify existing biases present in their training data, a significant challenge widely acknowledged [11]. Future work should explore the integration of multiple data sources and the implementation of algorithmic bias-auditing techniques.
- The Risk of "Hallucinations" and Factual Inaccuracy: Large Language Models are prone to generating plausible but incorrect information ("hallucinations"). While our process of summarizing existing abstracts mitigates this risk compared to open-ended generation, it is not entirely eliminated. As van Dis et al. (2023) [13] note, ensuring the factual accuracy of LLM outputs is a critical research priority. A future iteration of the platform could include a "human-in-the-loop" verification stage or automated cross-referencing to enhance reliability.
- The "Scaffold" Must Be Temporary: The ultimate goal of any pedagogical scaffold is to be removed once the learner is competent (Wood et al., 1976) [8]. There is a pedagogical risk that students might become overly reliant on the digest and neglect to engage with the primary literature themselves. Therefore, the successful implementation of our framework depends heavily on the educator, who must design tasks that explicitly require students to move beyond the digest and critically analyze the original sources.
- Lack of Empirical Validation: This paper presents a conceptual approach. While it is grounded in established learning theory, we have not yet conducted a formal, large-scale empirical study to measure its effectiveness. The next logical step is to design a controlled experiment with student groups to quantitatively assess the impact of AI SciBrief on the quality of research proposals, the time spent on literature review, and student confidence levels.

## VI. Conclusion and Future Work

### A. *Conclusion*

This paper introduced AI SciBrief, an LLM-powered platform, and proposed a pedagogical approach for its integration into higher education. Our central contribution is the positioning of this tool as a multi-level "gateway to research" that adapts to the evolving needs of students as they progress from undergraduate to doctoral studies.

We have demonstrated how this approach, grounded in established learning theories, can transform the research process: from simplifying topic discovery for undergraduates to accelerating landscape mapping for master's students and enabling interdisciplinary synthesis for doctoral candidates. Ultimately, we argue that AI SciBrief represents example a constructive model of human-AI collaboration in academia. It is not a replacement for the researcher but an empowerment tool - one that automates the laborious task of information filtering to free up the student's most valuable resources: their time, their critical thinking, and their capacity for creating new knowledge.

### B. *Future Work*

Building on the conceptual foundation laid out in this paper, our future work will proceed along three key tracks:

*1) Empirical Validation:* The immediate next step is to move from a conceptual framework to empirical validation. We plan to design and conduct a formal study with control and experimental student groups to quantitatively measure the impact of AI SciBrief. Key metrics will include the time spent on literature review, the quality of research proposals, and self-reported student confidence levels.

*2) Platform Enhancement and Trustworthiness:* To address the inherent limitations of current LLM technology, we will focus on enhancing the platform's reliability. This includes implementing an automated cross-referencing system to verify claims and minimize the risk of factual inaccuracies or "hallucinations." Furthermore, we will explore the integration of multiple scholarly databases beyond OpenAlex to mitigate potential data bias and broaden the scope of our analysis.

*3) Expansion of Disciplinary Coverage:* Our vision for AI SciBrief is to serve as a truly universal gateway to research. While our initial focus has been on Finance, Medicine, and Education, we are actively working to expand our coverage. Future development will involve adding new digests for a wider range of disciplines in the natural sciences, social sciences, and humanities, making the tool a comprehensive resource for the entire academic community.